# Hydrogenation and defect formation control the strength and ductility of MoS$_2$ nanosheets: Reactive molecular dynamics simulation

Mostafa Hasanian[1], Bohayra Mortazavi[2], Alireza Ostadhossein[*,3], Timon Rabczuk[2] and Adri C. T. van Duin[4]

[1]Department of Engineering Science and Mechanics, Penn State, University Park, Pennsylvania 16802, USA

[2]Institute of Structural Mechanics, Bauhaus-Universität Weimar, Marienstr. 15, D-99423 Weimar, Germany.

[3]Department of Engineering Science and Mechanics, Pennsylvania State University, 409A EES Building, University Park, USA.

[4]Department of Mechanical and Nuclear Engineering, Pennsylvania State University, University Park, USA

Two-dimensional (2D) molybdenum disulfide (MoS$_2$) has attracted significant attention because of its outstanding properties, suitable for application in several critical technologies like; solar cells, photocatalysis, lithium-ion batteries, nanoelectronics, and electrocatalysis. Similar to graphene and other 2D materials, the physical and chemical properties of MoS$_2$ can be tuned by the chemical functionalization and defects. In this investigation, our objective is to explore the mechanical properties of single-layer MoS$_2$ functionalized by the hydrogen atoms. We moreover analyze the effects of different types of defects on the mechanical response of MoS$_2$ at the room temperature. To investigate these systems, we conducted reactive molecular dynamics simulations using the ReaxFF forcefield. We demonstrate that an increase in the hydrogen adatoms or defects contents significantly affects the critical mechanical characteristics of MoS$_2$; elastic modulus, tensile strength, stretchability and failure behavior. Our reactive molecular dynamics results provide useful information concerning the mechanical response of hydrogenated and defective MoS$_2$ and the design of nanodevices.

## 1. Introduction

Transition metal dichalcogenides (TMD) are an emerging family of two dimensional (2D) materials that exhibit outstanding photoelectronic, thermal and mechanical properties. These attractive 2D nanostructures can be fabricated from bulk TMD structures by a top-down exfoliation approach [1], or they might be directly synthesized by a bottom-up technique such as molecular beam epitaxy [2] or chemical vapour deposition (CVD) [3]. Among the members of the TMD family, molybdenum disulphide ($MoS_2$) [4,5] can be considered as the forefront material [6], which has proven to yield very promising properties for diverse applications such as; solar cells, photocatalysis, lithium-ion batteries, nanoelectronics and electrocatalysis [5–11]. It is worthwhile to note its similarly to its counterpart graphene [12,13]; indeed bulk $MoS_2$ for several centuries has been also widely used as a lubricant [14]. An exciting fact about the $MoS_2$ and other members of TMD family is the polymorphism nature. In this regard, $MoS_2$ can exist in three different structural phases of 2H, 1T and 1T' [15,16], with contrasting electronic properties. The 2H phase of $MoS_2$ is its most stable structure, which is a well-known semiconductor, whereas the 1T configuration exhibits metallic electronic character. Interestingly, while a bulk $MoS_2$ structure shows an indirect band-gap, when it converts to a monolayer presents a direct band-gap semiconducting response, which is highly desirable for the application in nanoelectronics [4,17,18]. $MoS_2$ nanomembranes can show remarkable thermal stability and mechanical characteristics [19–22] with the elastic response comparable to that of the graphene [23,24].

The electronic and optical properties of mono/multi-layer $MoS_2$ can be further tuned through applying the mechanical strains [25,26]. Similar to graphene, chemical functionalization is also another very promising route for the engineering of the properties of $MoS_2$ nanomembranes [27–29]. Among different adatoms, hydrogenation of $MoS_2$ has been realized to efficiently tune its electronic and magnetic properties [30–32]. Experimentally synthesized 2H-$MoS_2$ nanomembranes are not completely perfect with only hexagonal atomic arrangements and the existence of different types of defects and atomic impurities are almost inevitable. According to the experimental observations, CVD grown $MoS_2$ nanosheets contain a wide variety of dislocation cores along the grain boundaries, such as; pentagon-heptagon, tetragon-tetragon, tetragon-hexagon, tetragon-octagon, and hexagon-octagon rings [33,34]. $MoS_2$ nanomembranes also show different types of points defects depending on the configuration of missing or substituting atoms, which have been experimentally realized by Zhou *et al.* [35]. Defects are among the most critical factors that can affect the mechanical properties of engineering materials and may easily lead to the structural failure [36–42]. In the case of 2D materials, defects can substantially alter the electronic, optical, chemical, thermal conduction and mechanical properties. Since the fabrication of defect-free materials is extremely challenging, they are commonly considered as an inevitable part of the material in the engineering designs. This way, studying the influence of defects on the various material properties has been a highly relevant research topic. On the other side, the possibility of tuning the properties of these materials through the control of type and population of defects is of great technological importance. While hydrogenated and defective $MoS_2$ have been focus of numerous experimental studies, only few first-principles and molecular dynamics studies are available concerning the mechanical properties of these systems [43–45]. Effect of point defects on the mechanical properties of $MoS_2$ have been studied using molecular dynamics simulations via various interatomic potentials [45–49]. However, in the previous studies only one or two defect types were considered, with more emphasis on S or Mo atoms vacancies [48,50]. The role of cracks with different orientations and the corresponding crack propagation mechanism have been investigated previously [51,52]. Effects of grain boundaries on the mechanical strength of $MoS_2$ has been also explored [47,53]. Despite of available theoretical and modelling efforts, a comprehensive study with respect to the effects of different point defect types that have been experimentally observed in $MoS_2$ has not been accomplished. Recently, the basal plane of $MoS_2$ nanosheet has been recognized as a promising non-precious catalyst for the hydrogen evolution reaction. Contrary to the first principle simulations [54,55] which commonly suits small scale simulations and neglect the thermal effects, MD simulations using ReaxFF reactive potential can provide fundamental insight into the role of hydrogen-adsorption on the stability and strength of $MoS_2$ catalysts. Thus, analysis of key mechanical properties of $MoS_2$ as a function of various coverage of hydrogen atoms and different defects contents play critical roles for the design of nanodevices exploiting the $MoS_2$ nanomembranes. In this work we performed systematic reactive molecular dynamics uniaxial tensile simulations to understand the mechanical responses of defective and hydrogenated $MoS_2$ nanosheets.

## 2. Molecular dynamics simulations

In the present investigation, mechanical responses of defective and hydrogenated $MoS_2$ were explored by performing uniaxial tensile simulations using the LAMMPS [56] package. OVITO software was also used to illustrate the atomistic structures [57]. We employed a recently developed ReaxFF [58] bond order potential for studying atomic-scale interactions of $MoS_2$+H systems. ReaxFF potentials have been applied to a wide range of systems, which allows the reactivity and describing the bond formation and breakage as well [59–62]. As discussed in our previous studies, the proposed ReaxFF by Ostadhossein *et al.* [58] yields accurate predictions for the mechanical properties [24] and thermal conductivity [63] of defect-free $MoS_2$.

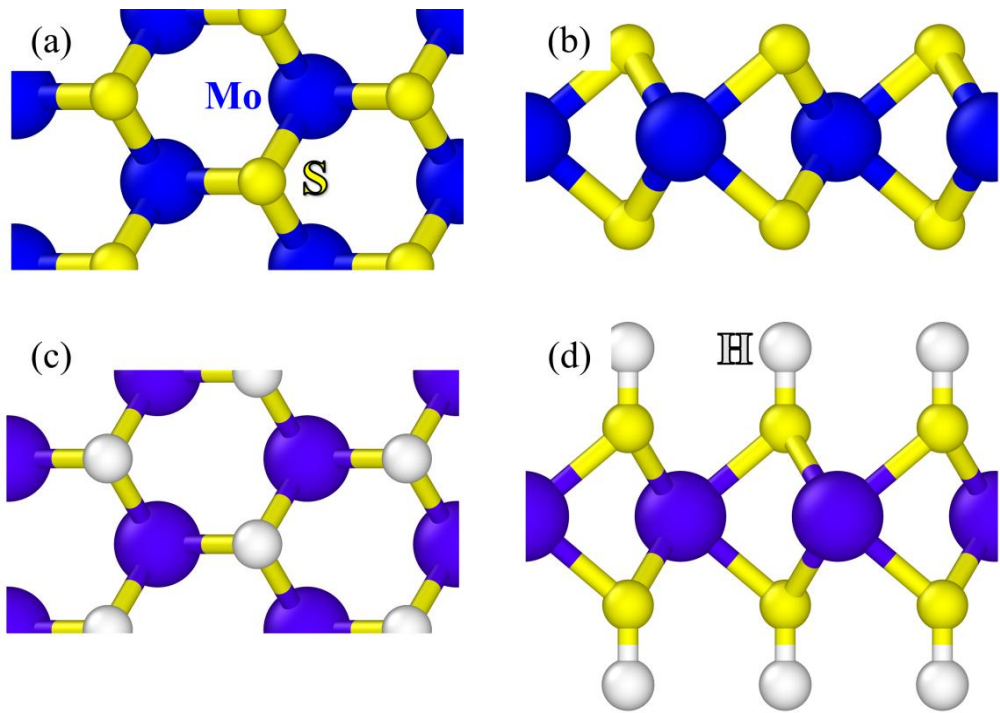

Fig. 1- Top and side views of pristine and hydrogenated single-layered 2H $MoS_2$.

For the evaluating the mechanical properties, we constructed an original atomistic model of pristine 2H-$MoS_2$ including 8280 individual atoms in a simulation box size with planar dimensions of 25 nm × 9.5 nm. To simulate $MoS_2$ sheets and not the nanoribbons, we applied periodic boundary condition in both planar directions. The time increment of all simulations was fixed at 0.25 fs. First, we conducted an energy minimization step using the conjugate gradient approach with termination criteria of $10^{-6}$ kcal/mol and $10^{-6}$ kcal/(mol.Å) for energy and force, respectively. Before applying the uniaxial tensile loading conditions, the structures were relaxed and equilibrated at the room temperature to ensure no residual stresses in the systems by performing the Nosé-Hoover barostat and thermostat (NPT) simulations with damping parameters of 62.5 fs and 625 fs for temperature and pressure, respectively.

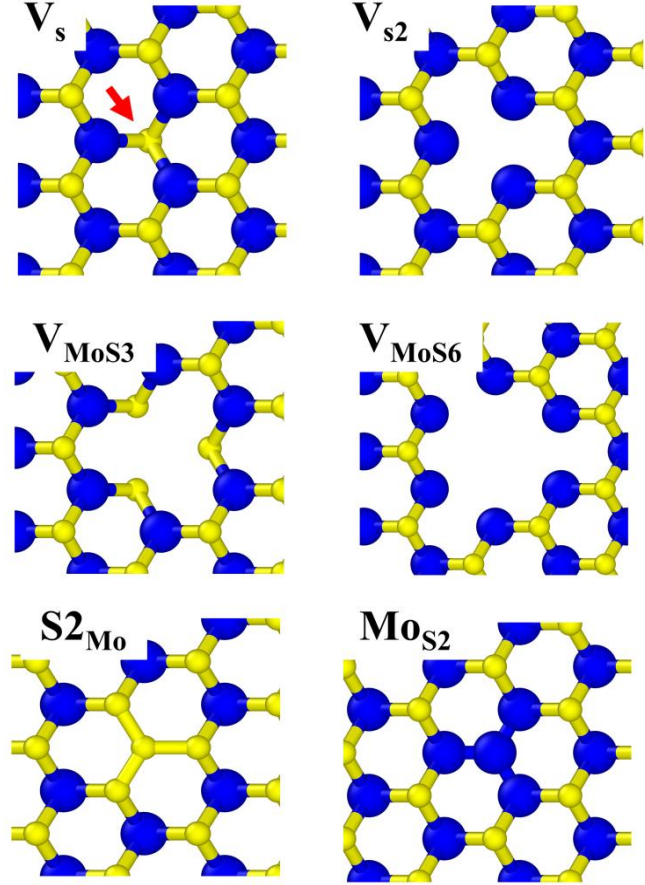

Fig. 2, Points defects in CVD grown $MoS_2$ as realized experimentally by Zhou *et al.* [35]; monosulfur vacancy ($V_S$), disulfur vacancy ($V_{S2}$), vacancy complex of Mo and connecting three sulfur ($V_{MoS3}$), vacancy complex of Mo and connecting three disulfur pairs ($V_{MoS6}$), and Mo atom substituting a $S_2$ pair column ($Mo_{S2}$) or a S2 column substituting a Mo atom ($S2_{Mo}$).

In order to apply the uniaxial loading condition, first, the periodic box size along the loading direction was increased at every simulation time step by a constant engineering strain rate of $1\times10^9\,s^{-1}$. To satisfy the uniaxial loading (stress) condition, stress along the two other directions perpendicular of the loading direction should remain negligible during the entire deformation process. Since the atoms are in contact with vacuum along the sheet normal direction, the stress along this direction is negligible. We therefore used NPT method to justify the periodic simulation box along the sample width to reach zero stress also in this direction and thus to accurately ensure uniaxial stress conditions. The Virial stresses at every simulation time step were calculated and averaged during every 250 fs intervals to report stress-strain curves. In the evaluation of stress values, the thickness of single-layer $MoS_2$ was assumed to be 6.1 Å based on the concept of bending rigidity [58]. For the fully hydrogenated $MoS_2$ ($MoS_2H_2$), the thickness was considered to be 8.9 Å, on the basis of thickness of pristine $MoS_2$ plus two H-S bond lengths (H-S bond length ≈1.4 Å [30]). The thickness of $MoS_2$ nanomembranes with various coverage of hydrogen atoms were calculated according to the content of hydrogen atoms (H%) as follows: 6.1+H%×2×1.4 (Å).

The atomic structure of pristine $MoS_2$ with 2H phase is illustrated in Fig. 1, which shows a hexagonal lattice with atomic stacking sequence of ABA. The fully hydrogenated $MoS_2$ is also depicted in Fig. 1. According to the experimental observations by Zhou *et al.* [35], points defects in 2H-$MoS_2$ nanomembranes include 6 major configurations; monosulfur vacancy ($V_S$), disulfur vacancy ($V_{S2}$), vacancy complex of Mo and connecting three sulfur ($V_{MoS3}$), vacancy complex of Mo and connecting three disulfur pairs ($V_{MoS6}$), and Mo atom substituting a $S_2$ pair column ($Mo_{S2}$) or a $S_2$ column substituting a Mo atom ($S2_{Mo}$). These 6 major defect types are illustrated in Fig. 2.

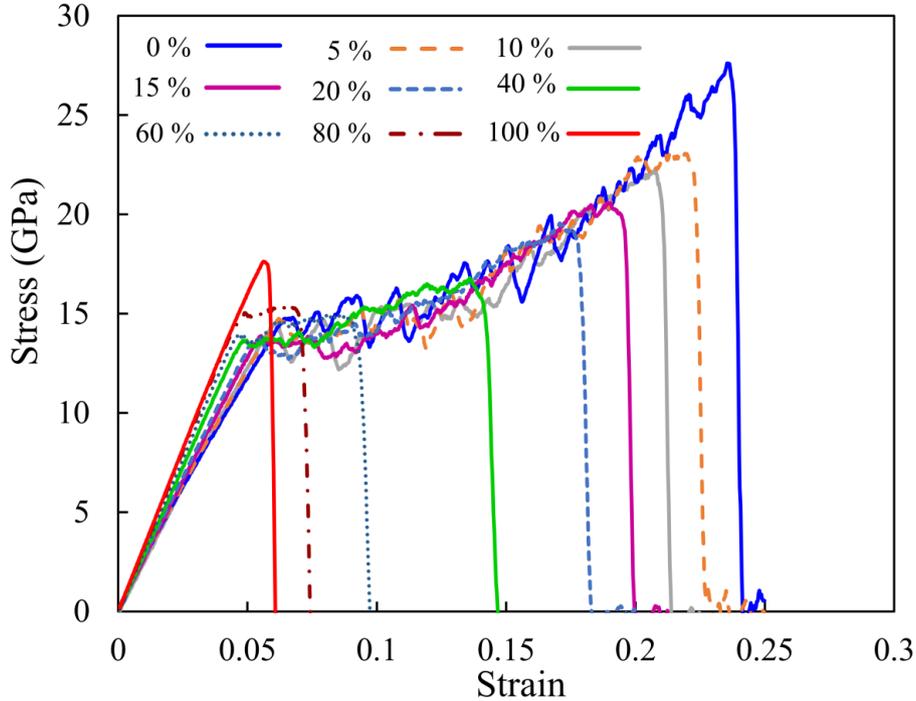

Fig. 3, Uniaxial stress-strain response of $MoS_2$ with different coverages of hydrogen atoms at the room temperature.

## 3. Results and discussions

The acquired stress-strain responses for defect-free and hydrogenated $MoS_2$ monolayers are shown in Fig. 3. For all structures, the stress-strain curves show initial linear relations which correspond to the elastic modulus. For the pristine $MoS_2$, after ending this first initial response a yield point is observable, after which the stress-strain curve exhibits an unusual and noisy pattern. As it was studied in our previous investigation [24], the pristine $MoS_2$ structure starts to undergo a phase transformation after this yield point. This phase transformation includes irreversible bond formations between S atoms on the same column, which helps the structure to flow easier along the loading direction. Fig. 4 compares the atomic structure of $MoS_2$ for pristine and fully hydrogenated samples at strain levels close to the rupture point. For fully hydrogenated $MoS_2$, S atoms almost keep their relative position with respect to Mo atoms, whereas pristine $MoS_2$ undergoes a transition by folding the S-Mo bond and forming S-S dimerization. Such a S-S dimerization can be explained by the bond-distance/ bond-order cut-off of S atoms in the ReaxFF potential. Interestingly, by increasing the hydrogen atoms content, the strain range associated with the phase-transformation decreases considerably and finally for the fully hydrogenated sample ($MoS_2H_2$) this initial yield point completely vanishes. This way, unusual and noisy stress patterns observable after the initial yield point in stress-strain curve corresponds to phase transition which occurs gradually during uniaxial loading. This shows that surface hydrogenation leads to the embrittlement of $MoS_2$ through preventing the S-S bond formation.

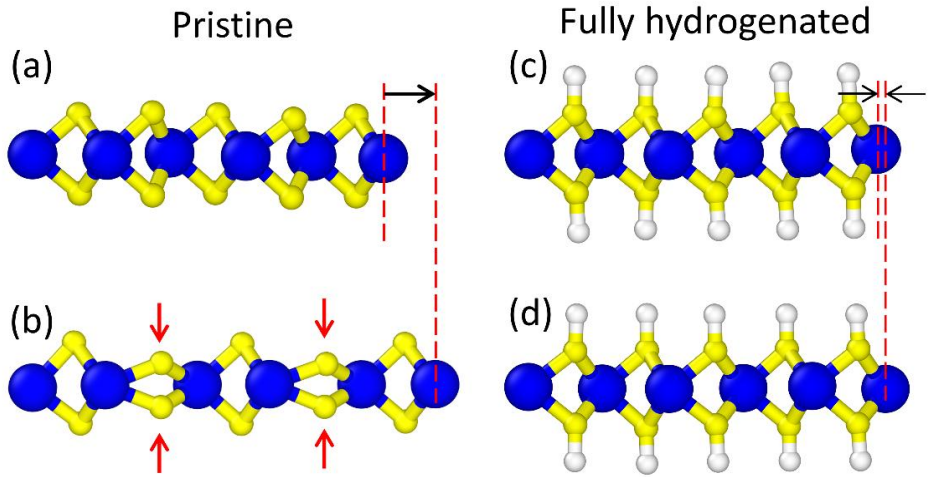

Fig. 4, Side views of pristine and fully hydrogenated $MoS_2$ (a and c) at stress-free condition and (b and d) before rupture at the tensile strength point. S-S dimerization is observable only in the case of pristine $MoS_2$.

The predicted results for mechanical properties of hydrogenated MoS$_2$ as a function of hydrogen atoms coverage are summarized in Fig. 5. As shown in Fig. 5a, by increasing the hydrogen atoms content the elastic modulus initially decreases by less than 1% and later after the 5% concentration of hydrogen atoms almost linearly increases. Nonetheless, for the hydrogen content of around 80%, a plateau in the elastic modulus can be seen, in which further increasing of hydrogen atoms only slightly enhance the elastic modulus. On the other side, the acquired results shown in Fig. 5b suggests that by increasing the hydrogen atoms concentration, the tensile strength initially drops sharply and later for the coverage of around 40% this drop reaches a plateau. Hydrogen adsorption therefore chemically weakens the Mo-S bonds in MoS$_2$ nanosheets and leads to the decrease of maximum tensile strength up to the 40% coverage of hydrogen atoms. In this case, the tensile strength of hydrogenated MoS$_2$ almost remains constant for the hydrogen contents in a range between 40-80%. Notably, the fully hydrogenated MoS$_2$ yields a higher tensile strength in comparison with the sample with 40% concentration of hydrogen atoms. Beyond the 80% coverage of hydrogen atoms, incorporation of further hydrogen atoms create a higher symmetry states, reducing the stress concentrations and finally resulting in a more uniform stress redistribution and enhancement of MoS$_2$ tensile strength.

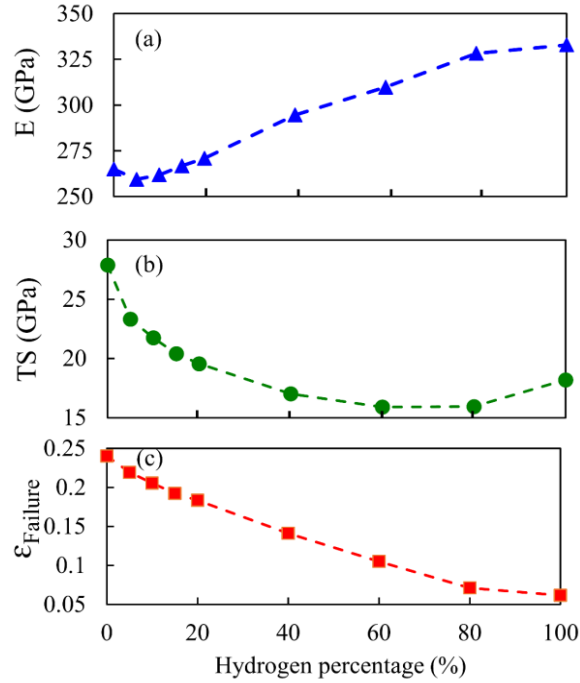

Fig. 5, Reactive molecular dynamics results for the elastic modulus (E), tensile strength (TS) and strain at failure ($\varepsilon_{Failure}$) of single-layer MoS$_2$ as a function of hydrogen atoms concentration at room temperature.

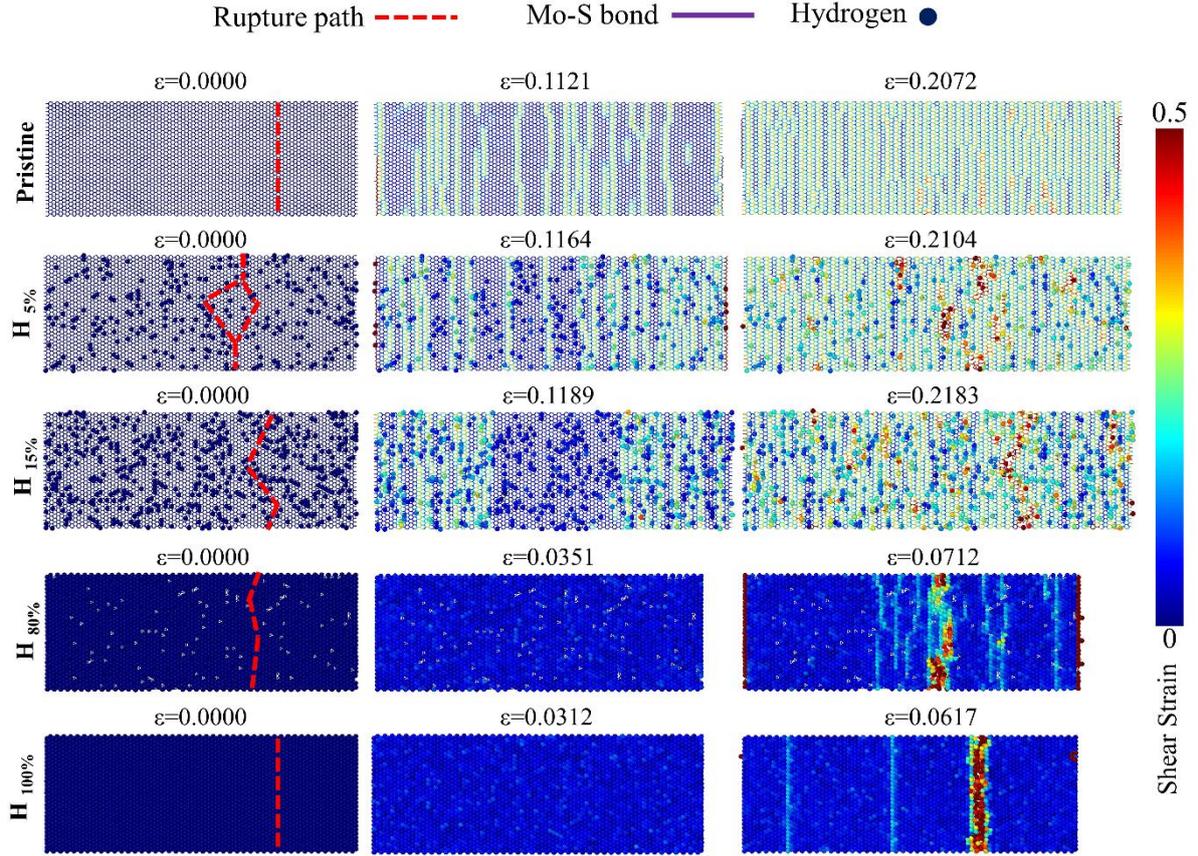

Fig. 6, Deformation process of single-layer $MoS_2$ with different content of hydrogen atoms at various uniaxial tensile strains (ε). Mo and S atoms are hidden and hydrogen atoms are depicted with exaggerated sizes, in order to better distinguish them. The contour illustrates the shear strain profile plotted using OVITO package [46].

To better understand the underlying mechanism with respect to the effects of hydrogen atoms on the failure process, in Fig. 6 shear strain contours for sample with various contents of hydrogen atoms and at different stain levels are compared. We remind that stretching of pristine $MoS_2$ monolayer beyond the yield point decreases the normal distance between sulfur atoms from the top ($S_{top}$) and bottom ($S_{top}$), leading to the consecutive covalent bonding of $S_{bot}$-$S_{top}$. S-S dimerizations result in shear strain localization, which are distinguishable by highlighted regions in Fig. 6 results. As it was discussed earlier, by increasing the hydrogen content, less phase transformation occur and such that rupture follow a path through zones with higher concentration of hydrogen atoms. This way, in the pristine and fully hydrogenated $MoS_2$ sheets cracking occur along almost straight lines perpendicular of the loading direction. In another word, symmetrical nature of pristine and fully hydrogenated $MoS_2$ sheets need more stresses for initiation of rupture, as in these cases the shear strain localization and subsequent stress concentrations cannot contribute to the failure. For the strain at failure point, our results depicted in Fig. 5c reveals a considerable decreasing trend by increasing the hydrogen atoms coverage. This observation can be also explained by the results illustrated in Fig. 6, in which total elongation at failure points reduce by increasing the hydrogen atoms coverage and formation of S-H covalent bonds. The formation of H-S bonds prevent the S-S dimerization and subsequently suppress the plastic deformation. This finding suggests that hydrogen atoms convert the deformation mechanism of $MoS_2$ from ductile to brittle in which the stretchability of fully hydrogenated $MoS_2$ is almost one-fifth of that of the pristine $MoS_2$ monolayer.

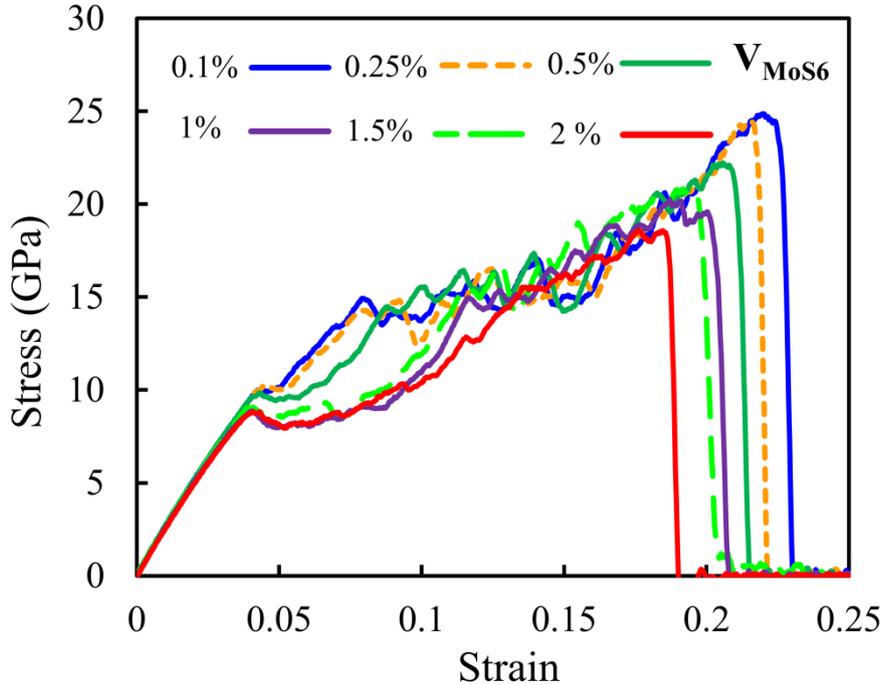

Fig. 7, Predicted uniaxial stress-strain response of $MoS_2$ with different concentration of $V_{MoS6}$ defects.

We next shift our attention to analyze the mechanical response of defective $MoS_2$. In Fig. 7 the predicted stress-strain responses for defective $MoS_2$ with different concentrations of $V_{MoS6}$ defects are illustrated. As expected, by increasing the defects concentration both the elastic modulus and tensile strength decrease. The suppression of elastic modulus is basically because of the decrease in the structure's rigidity as a result of bond removals. The decline in the tensile strength is originally due to the increase in the stress concentrations throughout the sample, which makes the rupture of the structure easier. For the $V_{MoS6}$ defect, by increasing the defect concentration, the strain at failure also decreases. Nevertheless, as it is clear yet the failure response is within the ductile range without a sharp rupture occurring after the initial elastic response, associated with S-S dimerization. By increasing the defects concentration, the strain at which the yield point occurs decreases, which consequently implies that the defects in the structure promote the phase transformation and corresponding S-S bonds formation.

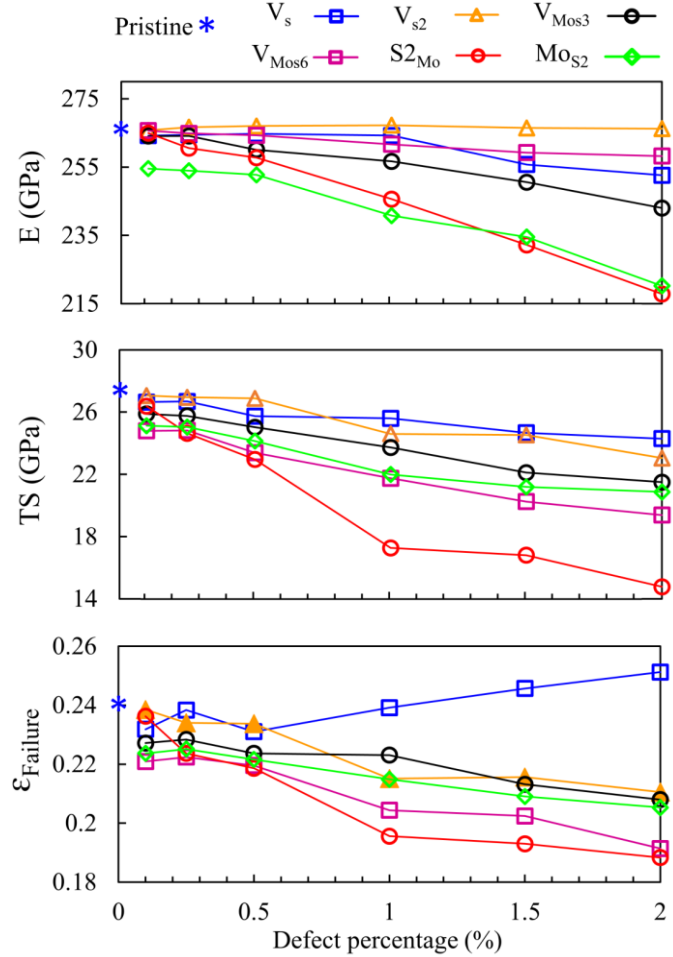

Fig. 8, Predicted elastic modulus (E), tensile strength (TS) and strain at failure ($\varepsilon_{Failure}$) of single-layer $MoS_2$ as a function of defects concentration.

The essential mechanical properties of defective $MoS_2$ as a function of different defects concentration are compared in Fig. 8. By increasing the defects concentrations the elastic modulus drops. Interestingly the elastic modulus shows the maximum senility for the structures with atomic substitutions rather than those with vacancies. The elastic modulus of single-layer $MoS_2$ with disulfur vacancies ($V_{S2}$ defect) decreases negligibly by increasing the defect concentration. With respect to the tensile strength, all defective structures show the declined performances in comparison with the pristine $MoS_2$. In this case, the sample with $S_2$ column substituting a Mo atom (S2$_{Mo}$ defect) indicates the maximum decrease in the tensile strength. The strain at failure for the defective MoS$_2$ structures also decreases by increasing the defects concentration, with an exception observable the case of monosulfur vacancy (V$_S$ defect) which exhibits an increasing trend.

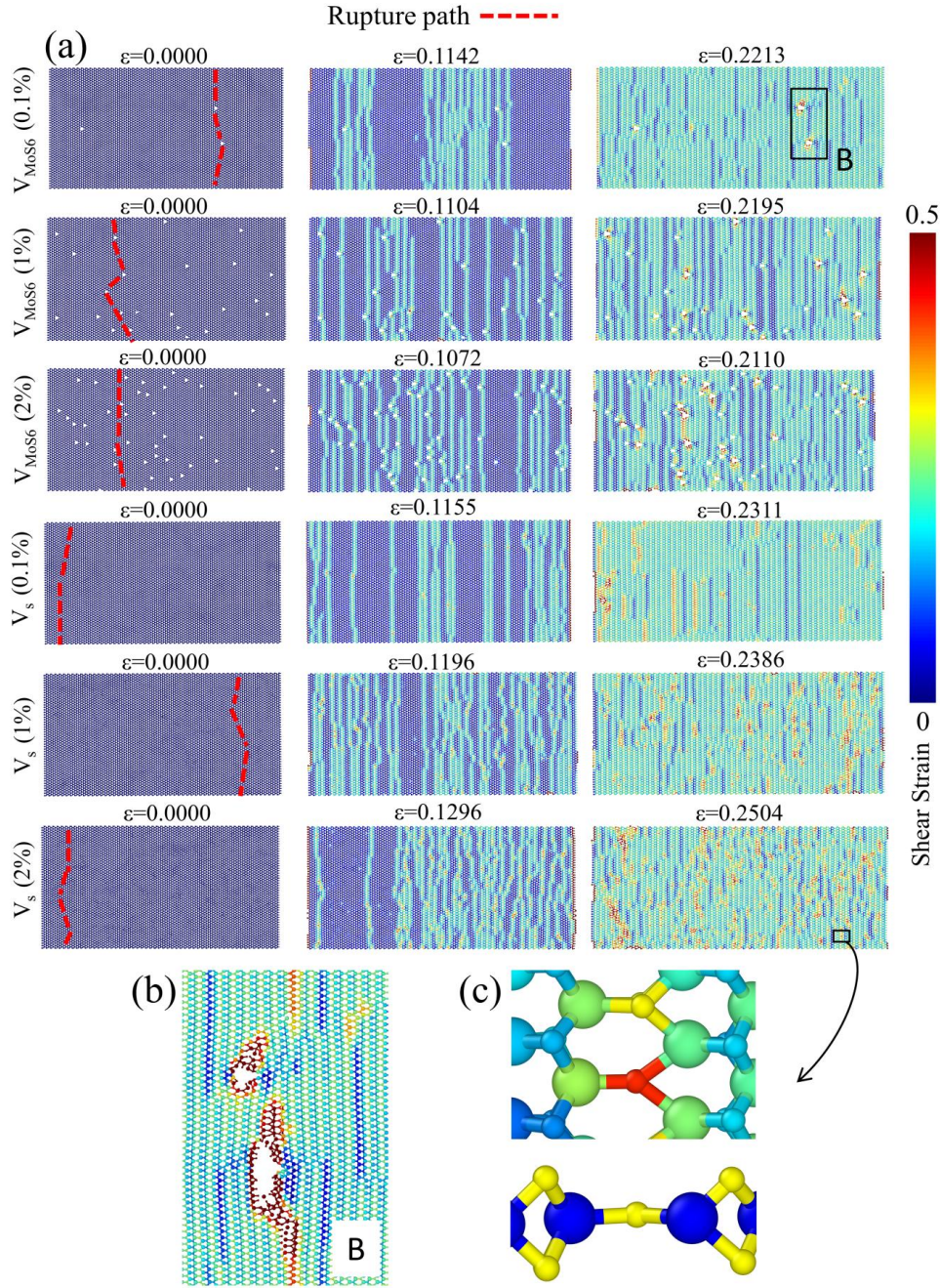

Fig. 9, Deformation process of single-layer MoS$_2$ with different types of point defects at various uniaxial tensile strains ($\varepsilon$). (b) shows zoomed view of defective MoS$_2$ with 0.1% content of V$_{MoS6}$ defects at $\varepsilon$=0.2213, which confirms high shear strains around the formed crack tips. (c) illustrates top and side zoomed views of defective MoS$_2$ with 2% content of V$_S$ defects at $\varepsilon$=0.2504, revealing the pivot-like behaviour in this system. Contours illustrate shear strain values.

In order to provide better insight with respect to the tensile responses of defective $MoS_2$, we analyzed the deformation process of these structures on the basis of shear strain contours in Fig. 9. In accordance with the results shown in Fig. 6, the highlighted areas with higher shear strain are associated with regions of phase transitions. Fig. 9a illustrates shear strain contours for $MoS_2$ sheet with 0.1% concentration of $V_{MoS6}$ defect type, which only includes three voids on the studied sheet. Interestingly, higher shear strains are observable around these defects, originated from stress concentrations existing around defects. To better understand this phenomena, Fig. 9b provides close look around two original point defects at the rupture strain, which clearly reveals that around the tip of propagating cracks due to the stress concentration the shear strains are maximum. In Fig. 9a we also studied the deformation process of two other $MoS_2$ nanosheets with different contents of $V_{MoS6}$ defects. In these cases around the original defects higher shear strains are clearly observable. These observations further confirm the remarkable role that the defects play in inducing stress concentrations, initiating the phase transformation and finally leading to the failure of the structure. The results shown in Fig. 9a show that by increasing the defects content shear strain field becomes more non-uniform which subsequently causes earlier initiation of rupture in the sample.

The obtained results shown in Fig. 8 illustrate an anomalous trend for $MoS_2$ monolayer with $V_s$ defects, in which in contrast with other types of defects, strain at failure increases by increasing the $V_s$ defects content. To better understand the mechanisms responsible for such an unexpected trend, we studied the tensile deformation of $MoS_2$ sheets with different concentration of $V_s$ defects in Fig. 9a. As it can be seen in the a zoomed view of Fig. 9c, lacking of single S atoms in the lattice results in the change in the deformation process such that the straining of S atoms in the absence of other S pairs act like pivots. These $MoS_2$ units can fold easily and form Mo-S-Mo planar atomic configurations. Such pivot-like behaviors can explain the enhanced stretchability of $MoS_2$ nanosheets with $V_s$ defects. We remind that as it was examined in our earlier work [24], S-S dimerizations are persistent upon the unloading of the $MoS_2$ structures which reveals the plastic deformation. Both the vacancy and substation defects in the $MoS_2$ lattice promote this phase-transformation at lower strain values and accordingly decrease the yield stress and its corresponding strain as well. This observation is also supported by the results shown in Fig. 9. Comparing to the presented results in Figs. 6 and 9, reveal lower density of regions with high shear strains in the hydrogenated MoS$_2$ in comparison with pristine and defective specimens, which highlights the impact of H-S bonds in embrittlement of the MoS$_2$ nanosheets."

4. Concluding remarks

We conducted ReaxFF based reactive molecular dynamics simulations to investigate the mechanical properties of defective and hydrogenated single-layer MoS$_2$. We studied the effects of randomly distributed hydrogen atoms and point defects (including both the atomic vacancies and substitutions) on the single-layer MoS$_2$ mechanical and failure response. For the pristine MoS$_2$, the stress-strain response exhibits an initial linear response followed by a yield point, after which the stress-strain curve shows a noisy pattern due to the S-S dimerizations and corresponding phase transformation. By increasing the hydrogen atoms content to 5% the elastic modulus initially decreases by around 2% and later almost linearly increases up to the 80% concentration of hydrogen atom, in which a plateau in the elastic modulus was observed. It was found that the hydrogen atoms adsorption weakens the Mo-S bonds in the MoS$_2$ nanosheets and thus decreases the maximum tensile strength up to the 80% coverage of hydrogen atoms. The fully hydrogenated MoS$_2$ nevertheless was found to yield a higher tensile strength in comparison with the sample with a lower concentration of hydrogen atoms because of providing more uniform load transfer conditions and suppression of stress-concentration. It was found that the H-S bonds prevent the S-S dimerization in the hydrogenated MoS$_2$, and therefore by increasing the hydrogen atoms coverage the stretchability decreases and MoS$_2$ failure mechanism change from ductile to completely brittle for the fully hydrogenated MoS$_2$. According to our atomistic results, it was predicted that by increasing the defects concentration the elastic modulus and tensile strength drop. Interestingly, the decline in the elastic modulus was found to be more considerable for the MoS$_2$ samples with atomic substitutions rather than those with vacancies. The stretchability of MoS$_2$ was predicted to generally decrease by increasing the defects concentration and only for the structures with mono sulfur vacancy the stretchability was observed to slightly enhance.

Acknowledgment

BM and TR greatly acknowledge the financial support by European Research Council for COMBAT project (Grant number 615132). AO and ACTvD acknowledge support from NSF grants DMR 1462980 and MIP/DMR-1539916.

**References**

(1) Coleman, J. N.; Lotya, M.; O'Neill, A.; Bergin, S. D.; King, P. J.; Khan, U.; Young, K.; Gaucher, A.; De, S.; Smith, R. J.; et al. Two-Dimensional Nanosheets Produced by Liquid Exfoliation of Layered Materials. *Science (80-. ).* **2011**, *331*, 568–571.

(2) Xenogiannopoulou, E.; Tsipas, P.; Aretouli, K. E.; Tsoutsou, D.; Giamini, S. A.; Bazioti, C.; Dimitrakopulos, G. P.; Komninou, P.; Brems, S.; Huyghebaert, C.; et al. High-Quality, Large-Area MoSe2 and MoSe2/Bi2Se3 Heterostructures on AlN(0001)/Si(111) Substrates by Molecular Beam Epitaxy. *Nanoscale* **2015**, *7*, 7896–7905.

(3) Lee, Y. H.; Zhang, X. Q.; Zhang, W.; Chang, M. T.; Lin, C. T.; Chang, K. D.; Yu, Y. C.; Wang, J. T.; Chang, C. S.; Li, L. J.; et al. Synthesis of Large-Area MoS2 Atomic Layers with Chemical Vapor Deposition. *Adv Mater* **2012**, *24*, 2320–2325.

(4) Radisavljevic, B.; Radenovic, A.; Brivio, J.; Giacometti, V.; Kis, A. Single-Layer MoS2 Transistors. *Nat. Nanotechnol.* **2011**, *6*, 147–150.

(5) Wang, Q. H.; Kalantar-Zadeh, K.; Kis, A.; Coleman, J. N.; Strano, M. S. Electronics and Optoelectronics of Two-Dimensional Transition Metal Dichalcogenides. *Nat. Nanotechnol.* **2012**, *7*, 699–712.

(6) Chua, X. J.; Tan, S. M.; Chia, X.; Sofer, Z.; Luxa, J.; Pumera, M. The Origin of MoS$_2$ Significantly Influences Its Performance for the Hydrogen Evolution Reaction Due to Differences in Phase Purity. *Chem. - A Eur. J.* **2017**, *23*, 3169–3177.

(7) Li, S.; Wang, S.; Tang, D. M.; Zhao, W.; Xu, H.; Chu, L.; Bando, Y.; Golberg, D.; Eda, G. Halide-Assisted Atmospheric Pressure Growth of Large WSe2 and WS2 Monolayer Crystals. *Appl. Mater. Today* **2015**, *1*, 60–66.

(8) Ambrosi, A.; Sofer, Z.; Pumera, M. Lithium Intercalation Compound Dramatically Influences the Electrochemical Properties of Exfoliated MoS$_2$. *Small* **2015**, *11*, 605–612.

(9) Ambrosi, A.; Pumera, M. Templated Electrochemical Fabrication of Hollow Molybdenum Sulfide Microstructures and Nanostructures with Catalytic Properties for Hydrogen Production. *ACS Catal.* **2016**, *6*, 3985–3993.

(10) Eftekhari, A. Molybdenum Diselenide (MoSe2) for Energy Storage, Catalysis, and Optoelectronics. *Appl. Mater. Today* **2017**, *8*, 1–17.

(11) Guha, R.; Mohajerani, F.; Collins, M.; Ghosh, S.; Sen, A.; Velegol, D. Chemotaxis of Molecular Dyes in Polymer Gradients in Solution. *J. Am. Chem. Soc.* **2017**, *139*, 15588–15591.

(12) Novoselov, K. S.; Geim, A. K.; Morozov, S. V; Jiang, D.; Zhang, Y.; Dubonos, S. V; Grigorieva, I. V; Firsov, A. A. Electric Field Effect in Atomically Thin Carbon Films. *Science* **2004**, *306*, 666–669.

(13) Geim, A. K.; Novoselov, K. S. The Rise of Graphene. *Nat. Mater.* **2007**, *6*, 183–191.

(14) Sliney, H. E. Solid Lubricant Materials for High Temperatures - a Review. *Tribol. Int.* **1982**, *15*, 303–315.

(15) Chhowalla, M.; Voiry, D.; Yang, J.; Shin, H. S.; Loh, K. P. Phase-Engineered Transition-Metal Dichalcogenides for Energy and Electronics. *MRS Bull.* **2015**, *40*, 585–591.

(16) Calandra, M. Chemically Exfoliated Single-Layer MoS 2: Stability, Lattice Dynamics, and Catalytic Adsorption from First Principles. *Phys. Rev. B - Condens. Matter Mater. Phys.* **2013**, *88*.

(17) Splendiani, A.; Sun, L.; Zhang, Y.; Li, T.; Kim, J.; Chim, C. Y.; Galli, G.; Wang, F. Emerging Photoluminescence in Monolayer MoS2. *Nano Lett.* **2010**, *10*, 1271–1275.

(18) Lopez-Sanchez, O.; Lembke, D.; Kayci, M.; Radenovic, A.; Kis, A. Ultrasensitive Photodetectors Based on Monolayer MoS2. *Nat. Nanotechnol.* **2013**, *8*, 497–501.

(19) Bertolazzi, S.; Brivio, J.; Kis, A. Stretching and Breaking of Ultrathin MoS 2. *ACS Nano* **2011**, *5*, 9703–9709.

(20) Castellanos-Gomez, A.; Poot, M.; Steele, G. a.; van der Zant, H. S. J.; Agrait, N.; Rubio-Bollinger, G. Mechanical Properties of Freely Suspended Semiconducting Graphene-like Layers Based on MoS2. *Nanoscale Res. Lett.* **2012**, *7*, 233.

(21) Feldman, J. L. Elastic Constants of 2H-MoS2 and 2H-NbSe2 Extracted from Measured Dispersion Curves and Linear Compressibilities. *J. Phys. Chem. Solids* **1976**, *37*, 1141–1144.

(22) Farahani, H.; Rajabpour, A.; Khanaki, M.; Reyhani, A. Interfacial Thermal Resistance between Few-Layer MoS2 and Silica Substrates: A Molecular Dynamics Study. *Comput. Mater. Sci.* **2018**, *142*, 1–6.

(23) Li, J.; Medhekar, N. V.; Shenoy, V. B. Bonding Charge Density and Ultimate Strength of Monolayer Transition Metal Dichalcogenides. *J. Phys. Chem. C* **2013**, *117*, 15842–15848.

(24) Mortazavi, B.; Ostadhossein, A.; Rabczuk, T.; van Duin, A. C. T. Mechanical Response of All-MoS2 Single-Layer Heterostructures: A ReaxFF Investigation. *Phys. Chem. Chem. Phys.* **2016**, *18*, 23695–23701.

(25) Conley, H. J.; Wang, B.; Ziegler, J. I.; Haglund, R. F.; Pantelides, S. T.; Bolotin, K. I. Bandgap Engineering of Strained Monolayer and Bilayer MoS2. *Nano Lett.* **2013**, *13*, 3626–3630.

(26) Castellanos-Gomez, A.; Roldán, R.; Cappelluti, E.; Buscema, M.; Guinea, F.; Van Der Zant, H. S. J.; Steele, G. A. Local Strain Engineering in Atomically Thin MoS2. *Nano Lett.* **2013**, *13*, 5361–5366.

(27) Voiry, D.; Goswami, A.; Kappera, R.; Silva, C. D. C. C. E.; Kaplan, D.; Fujita, T.; Chen, M.; Asefa, T.; Chhowalla, M. Covalent Functionalization of Monolayered Transition Metal Dichalcogenides by Phase Engineering. *Nat. Chem.* **2015**, *7*, 45–49.

(28) Backes, C.; Berner, N. C.; Chen, X.; Lafargue, P.; LaPlace, P.; Freeley, M.; Duesberg, G. S.; Coleman, J. N.; McDonald, A. R. Functionalization of Liquid-Exfoliated Two-Dimensional 2H-MoS2. *Angew. Chemie - Int. Ed.* **2015**, *54*, 2638–2642.

(29) Chen, X.; Berner, N. C.; Backes, C.; Duesberg, G. S.; McDonald, A. R. Functionalization of Two-Dimensional MoS2: On the Reaction between MoS2 and Organic Thiols. *Angew. Chemie - Int. Ed.* **2016**, *55*, 5803–5808.

(30) Shi, H.; Pan, H.; Zhang, Y. W.; Yakobson, B. I. Strong Ferromagnetism in Hydrogenated Monolayer MoS2 Tuned by Strain. *Phys. Rev. B - Condens. Matter Mater. Phys.* **2013**, *88*.

(31) Ouyang, F.; Yang, Z.; Ni, X.; Wu, N.; Chen, Y.; Xiong, X. Hydrogenation-Induced Edge Magnetization in Armchair MoS2 Nanoribbon and Electric Field Effects. *Appl. Phys. Lett.* **2014**, *104*, 71901.

(32) Zhang, W.; Zhang, Z. J.; Yang, W. Stability and Electronic Properties of Hydrogenated MoS2 Mono Layer: A First-Principles Study. *J. Nanosci. Nanotechnol.* **2015**, *15*, 8075–8080.

(33) van der Zande, A. M.; Huang, P. Y.; Chenet, D. a; Berkelbach, T. C.; You, Y.; Lee, G.-H.; Heinz, T. F.; Reichman, D. R.; Muller, D. a; Hone, J. C. Grains and Grain Boundaries in Highly Crystalline Monolayer Molybdenum Disulphide. *Nat. Mater.* **2013**, *12*, 554–561.

(34) Najmaei, S.; Liu, Z.; Zhou, W.; Zou, X.; Shi, G.; Lei, S.; Yakobson, B. I.; Idrobo, J.-C.; Ajayan, P. M.; Lou, J. Vapour Phase Growth and Grain Boundary Structure of Molybdenum Disulphide Atomic Layers. *Nat. Mater.* **2013**, *12*, 754–759.

(35) Zhou, W.; Zou, X.; Najmaei, S.; Liu, Z.; Shi, Y.; Kong, J.; Lou, J.; Ajayan, P. M.; Yakobson, B. I.; Idrobo, J. C. Intrinsic Structural Defects in Monolayer Molybdenum Disulfide. *Nano Lett.* **2013**, *13*, 2615–2622.

(36) Yang, H.; Huang, X.; Liang, W.; Van Duin, A. C. T.; Raju, M.; Zhang, S. Self-Weakening in Lithiated Graphene Electrodes. *Chem. Phys. Lett.* **2013**, *563*, 58–62.

(37) Terdalkar, S. S.; Huang, S.; Yuan, H.; Rencis, J. J.; Zhu, T.; Zhang, S. Nanoscale Fracture in Graphene. *Chem. Phys. Lett.* **2010**, *494*, 218–222.

(38) Huang, X.; Yang, H.; Van Duin, A. C. T.; Hsia, K. J.; Zhang, S. Chemomechanics Control of Tearing Paths in Graphene. *Phys. Rev. B - Condens. Matter Mater. Phys.* **2012**, *85*.

(39) Abadi, R.; Uma, R. P.; Izadifar, M.; Rabczuk, T. The Effect of Temperature and Topological Defects on Fracture Strength of Grain Boundaries in Single-Layer Polycrystalline Boron-Nitride Nanosheet. *Comput. Mater. Sci.* **2016**, *123*, 277–286.

(40) Abadi, R.; Uma, R. P.; Izadifar, M.; Rabczuk, T. Investigation of Crack Propagation and Existing Notch on the Mechanical Response of Polycrystalline Hexagonal Boron-Nitride Nanosheets. *Comput. Mater. Sci.* **2017**, *131*, 86–99.

(41) Shirazi, A. H. N.; Abadi, R.; Izadifar, M.; Alajlan, N.; Rabczuk, T. Mechanical Responses of Pristine and Defective C3N Nanosheets Studied by Molecular Dynamics Simulations. *Comput. Mater. Sci.* **2018**, *147*, 316–321.

(42) Sadeghzadeh, S. Borophene Sheets with In-Plane Chain-like Boundaries; a Reactive Molecular Dynamics Study. *Comput. Mater. Sci.* **2018**, *143*, 1–14.

(43) Wang, W.; Yang, C.; Bai, L.; Li, M.; Li, W. First-Principles Study on the Structural and Electronic Properties of Monolayer MoS2 with S-Vacancy under Uniaxial Tensile Strain. *Nanomaterials* **2018**, *8*, 74.

(44) Jiang, J.-W.; Park, H. S.; Rabczuk, T. Molecular Dynamics Simulations of Single-Layer Molybdenum Disulphide (MoS2): Stillinger-Weber

Parametrization, Mechanical Properties, and Thermal Conductivity. *J. Appl. Phys.* **2013**, *114*, 064307.

(45) Hong, J.; Wang, Y.; Wang, A.; Lv, D.; Jin, C.; Xu, Z.; Probert, M. I. J.; Yuan, J.; Zhang, Z. Atomistic Dynamics of Sulfur-Deficient High-Symmetry Grain Boundaries in Molybdenum Disulfide. *Nanoscale* **2017**, *9*, 10312–10320.

(46) Haldar, S.; Vovusha, H.; Yadav, M. K.; Eriksson, O.; Sanyal, B. Systematic Study of Structural, Electronic, and Optical Properties of Atomic-Scale Defects in the Two-Dimensional Transition Metal Dichalcogenides MX2 (M=Mo, W; X=S, Se, Te). *Phys. Rev. B* **2015**, *92*, 235408.

(47) Dang, K. Q.; Spearot, D. E. Effect of Point and Grain Boundary Defects on the Mechanical Behavior of Monolayer MoS2 under Tension via Atomistic Simulations. *J. Appl. Phys.* **2014**, *116*.

(48) Li, M.; Wan, Y.; Wang, W. Prediction of Mechanical Properties for Defective Monolayer MoS2 with Single Molybdenum Vacancy Defects Using Molecular Dynamics Simulations. In *2017 IEEE 17th International Conference on Nanotechnology (IEEE-NANO)*; IEEE, 2017; pp. 9–12.

(49) Xiong, S.; Cao, G. Molecular Dynamics Simulations of Mechanical Properties of Monolayer MoS2. *Nanotechnology* **2015**, *26*, 185705.

(50) Li, M.; Wan, Y.; Tu, L.; Yang, Y.; Lou, J. The Effect of VMoS3 Point Defect on the Elastic Properties of Monolayer MoS2 with REBO Potentials. *Nanoscale Res. Lett.* **2016**, *11*, 155.

(51) Wang, X.; Tabarraei, A.; Spearot, D. E. Fracture Mechanics of Monolayer Molybdenum Disulfide. *Nanotechnology* **2015**, *26*, 175703.

(52) Bao, H.; Huang, Y.; Yang, Z.; Sun, Y.; Bai, Y.; Miao, Y.; Chu, P. K.; Xu, K.; Ma, F. Molecular Dynamics Simulation of Nanocrack Propagation in Single-Layer MoS2 Nanosheets. *J. Phys. Chem. C* **2018**, *122*, 1351–1360.

(53) Wu, J.; Cao, P.; Zhang, Z.; Ning, F.; Zheng, S.; He, J.; Zhang, Z. Grain-Size-Controlled Mechanical Properties of Polycrystalline Monolayer MoS2. *Nano Lett.* **2018**, *18*, 1543–1552.

(54) Xu, Y.; Li, Y.; Chen, X.; Zhang, C.; Zhang, R.; Lu, P. First-Principle Study of Hydrogenation on Monolayer MoS2. *AIP Adv.* **2016**, *6*, 075001.

(55) Keong Koh, E. W.; Chiu, C. H.; Lim, Y. K.; Zhang, Y.-W.; Pan, H. Hydrogen Adsorption on and Diffusion through MoS2 Monolayer: First-Principles Study. *Int. J. Hydrogen Energy* **2012**, *37*, 14323–14328.

(56) Plimpton, S. Fast Parallel Algorithms for Short-Range Molecular Dynamics. *J. Comput. Phys.* **1995**, *117*, 1–19.

(57) Stukowski, A. Visualization and Analysis of Atomistic Simulation Data with OVITO–the Open Visualization Tool. *Model. Simul. Mater. Sci. Eng.* **2009**, *18*, 015012.

(58) Ostadhossein, A.; Rahnamoun, A.; Wang, Y.; Zhao, P.; Zhang, S.; Crespi, V. H.; van Duin, A. C. T. ReaxFF Reactive Force-Field Study of Molybdenum Disulfide (MoS2). *J. Phys. Chem. Lett.* **2017**, *8*, 631–640.

(59) Islam, M. M.; Ostadhossein, A.; Borodin, O.; Yeates, A. T.; Tipton, W. W.; Hennig, R. G.; Kumar, N.; van Duin, A. C. T. ReaxFF Molecular Dynamics Simulations on Lithiated Sulfur Cathode Materials. *Phys. Chem. Chem. Phys.* **2015**, *17*, 3383–3393.

(60) Yoon, K.; Ostadhossein, A.; Van Duin, A. C. T. Atomistic-Scale Simulations of the Chemomechanical Behavior of Graphene under Nanoprojectile Impact. *Carbon N. Y.* **2016**, *99*, 58–64.

(61) Ostadhossein, A.; Cubuk, E. D.; Tritsaris, G. A.; Kaxiras, E.; Zhang, S.; van Duin, A. C. T. Stress Effects on the Initial Lithiation of Crystalline Silicon Nanowires: Reactive Molecular Dynamics Simulations Using ReaxFF. *Phys. Chem. Chem. Phys.* **2015**, *17*, 3832–3840.

(62) Osti, N. C.; Naguib, M.; Ostadhossein, A.; Xie, Y.; Kent, P. R. C.; Dyatkin, B.; Rother, G.; Heller, W. T.; van Duin, A. C. T.; Gogotsi, Y.; *et al.* Effect of Metal Ion Intercalation on the Structure of MXene and Water Dynamics on Its Internal Surfaces. *ACS Appl. Mater. Interfaces* **2016**, *8*, 8859–8863.

(63) Mortazavi, B.; Quey, R.; Ostadhossein, A.; Villani, A.; Moulin, N.; van Duin, A. C. T.; Rabczuk, T. Strong Thermal Transport along Polycrystalline Transition Metal Dichalcogenides Revealed by Multiscale Modeling for MoS 2. *Appl. Mater. Today* **2017**, *7*, 67–76.